# Crystallographic phase transition and high-$T_c$ superconductivity in LaFeAsO:F

T Nomura[1], S W Kim[2], Y Kamihara[3], M Hirano[3], P V Sushko[4,5], K Kato[6], M Takata[6], A L Shluger[4] and H Hosono[1,3,7]*

[1] Materials and Structures Laboratory, Tokyo Institute of Technology, Yokohama 226-8503, Japan
[2] Secure Materials Center, Materials and Structures Laboratory, Tokyo Institute of Technology, Yokohama 226-8503, Japan
[3] ERATO-SORST, JST, in Frontier Research Center, Tokyo Institute of Technology, Yokohama 226-8503, Japan
[4] London Center for Nanotechnology & Department of Physics and Astronomy, University of College London, Gower Street, London WC1E 6BT, UK
[5] WPI Advanced Institute for Materials Research, Tohoku University, Sendai 980-8577, Japan
[6] RIKEN SPring-8 Center, Sayo-gun, Hyogo, 679-5148, Japan
[7] Frontier Research Center, Tokyo Institute of Technology, Yokohama 226-8503, Japan

Corresponding E-mail: hosono@msl.titech.ac.jp

**Abstract.** Undoped LaFeAsO, parent compound of the newly found high-$T_c$ superconductor, exhibits a sharp decrease in the temperature-dependent resistivity at ~160 K. The anomaly can be suppressed by F doping and the superconductivity appears correspondingly, suggesting a close associate of the anomaly with the superconductivity. We examined the crystal structures, magnetic properties and superconductivity of undoped (normal conductor) and 14 at.% F-doped LaFeAsO ($T_c$ = 20 K) by synchrotron X-ray diffraction, DC magnetic measurements, and *ab initio* calculations to demonstrate that the anomaly is associated with a phase transition from tetragonal (*P*4/*nmm*) to orthorhombic (*Cmma*) phases at ~160 K as well as an antiferromagnetic transition at ~140 K. These transitions can be explained by spin configuration-dependent potential energy surfaces derived from the *ab initio* calculations. The suppression of the transitions is ascribed to interrelated effects of geometric and electronic structural changes due to doping by F⁻ ions.
Keyword: Superconductivity, Oxypnictide, Phase transition,

## 1. Introduction

Discovery of high-$T_c$ superconductor LaFeAsO:F ($T_c$ = 26 K) [1] has caused a recurrence of a new superconductivity boom similar to that caused by the finding of layered copper oxides. With either applying the external pressure of ~3 GPa [2] or replacing La with other rare-earth-metal elements such as Sm [3], it has been possible to achieve $T_c$ beyond 55 K. The undoped LaFeAsO, which is the parent compound to the superconductor, is a member of the large $LnT_M Pn$O family, where $Ln$ represents a 4$f$ rare earth element, $T_M$ – a transition metal element with a more than half-filled 3$d$ shell, and $Pn$ – a pnicogen element [4–8]. They have a common ZrCuSiAs-type crystal structure, belonging to the tetragonal $P4/nmm$ space group (figure 1(a)). The crystal is formed by an alternating stack of electrically charged $Ln$O and $T_M Pn$ layers, and can be represented also as $(LnO)^{+\delta}(T_M Pn)^{-\delta}$. Several high-$T_c$ superconductors have been also discovered in other Fe based analogous compounds including $A$Fe$_2$As$_2$ ($A$ is an alkali earth metal element such as Sr and Ba) [9,10] and LiFeAs [11,12]. This, together with the observed lower $T_c$ in LaFePO [13] and LaNiP(As)O ($T_c$ = 2–4 K) [14–16], strongly suggests that FeAs layers play a leading role in the appearance of the high $T_c$. This view is supported in part by theoretical analysis of the energy band structure based on the density functional theory, revealing that five Fe 3$d$ orbitals hybridized with As 4$p$ contribute to the Fermi surface whereas LaO, $A$ and Li layers are blocking or spacer layers which act as a charge reservoir.

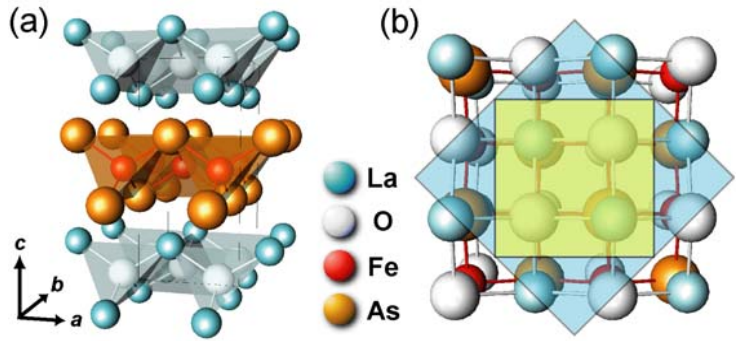

**Figure 1.** Crystal structure of LaFeAsO. (a) Schematic view of crystal structure demonstrates the layered structure. Distorted tetrahedrons of FeAs$_4$ are connected in an edge sharing manner to form the FeAs layer. (b) Top view of the crystal structure from $c$-direction. The inner square represents the unit cell in the tetragonal phase ($P4/nmm$). The outer square is that in the orthorhombic phase ($Cmma$). The unit cell of the orthorhombic phase rotates by 45° from that of the tetragonal phase and the lattice constants expands by √2, resulting in the increase in number of the chemical formula in a unit cell from 2 to 4.

With changing $T_M$ element, electronic and magnetic properties of the $LnT_MPnO$ compounds vary such that it is an antiferromagnetic insulator for $T_M$ = Mn [17], a superconductor for $T_M$ = Fe [1,13,18], a ferromagnetic metal for $T_M$ = Co [19], and again a superconductor for $T_M$ = Ni [14–16]. Further, the replacement of La with other rare earth elements having 4$f$ spin magnetic moments provides an additional magnetic interaction and facilitates antiferromagnetic ordering of spins at low temperatures, which results in coexistence of the superconductivity with antiferromagnetism for $Ln$ = Ce, Nd, Pr and Sm, for instance, in ref. 20. More importantly, the substitution of the rare earth elements with smaller ionic radii makes lattice constants smaller and enables raising $T_c$ up to 56 K [21,22].

High-$T_c$ superconductivity in $Ln$FeAsO materials is realized upon doping with electrons; this can be achieved either by replacing $O^{2-}$ ions in the reservoir layers by $F^-$ ions or by forming oxygen vacancies [23,24]. The doping can also be achieved by partially substituting Co for Fe in the FeAs layer [25,26]. On the contrary, $A$Fe$_2$As$_2$ compounds can undergo the superconducting transition only after hole doping, which is achieved through the replacement of $A$ elements with potassium.

Parent compounds to the high-$T_c$ superconductors show a rapid decrease in their electrical resistivity ($\rho$), which is clearly seen on the resistivity-temperature ($T$) curves with a kink at ~160 K ($T_{anom}$). This anomaly has been attributed to the combined effect of a crystallographic phase transition at ~160 K, and an antiferromagnetic ordering of the Fe spins at a slightly lower temperature of ~140 K [27–31]. Both transitions can be simultaneously suppressed by the electron or hole doping, suggesting a close association of these phase transitions with the superconductivity observed in the doped compounds.

The Fe-based and the Cu-based superconductors have a common feature in that superconductivity is attained by providing itinerant electron or hole carriers to the two-dimensional transport layers containing 3$d$ transition metal elements. However, they differ distinctly from each other in that nine 3$d$ electrons (one hole) are involved for $Cu^{2+}$, which forms ionic bond with oxide ions, whereas six 3$d$ electrons participate in a more complex interplay of Fe–Fe and Fe–As bonding.

In this study, we examine the crystal structures, magnetic properties and superconductivity of undoped and 14 at.% F-doped LaFeAsO ($T_c$ = 20 K) by Rietveld refinement of synchrotron X-ray diffraction, DC magnetic measurements, and *ab initio* calculations. We demonstrate that the undoped LaFeAsO undergoes a phase transition from tetragonal (*P*4/*nmm*) to orthorhombic (*Cmma*) phases at ~160 K as well as an antiferromagnetic transition at ~140 K. These transitions can be explained by spin configuration-dependent potential energy surfaces derived from the *ab initio* calculations. Doping by $F^-$ ions in the LaO layers suppresses both transitions, which is ascribed to interrelated effects of geometric and electronic structural changes. Our results demonstrate how doping of electrons in the FeAs layer can increase the $T_c$, suggesting that the interplay between the charge and spin density fluctuations is responsible for the high $T_c$ in LaFeAsO:F.

## 2. Experimental and *ab initio* calculation procedures

Undoped and 14% F-doped LaFeAsO samples employed in this study were prepared and the content of the F dopant in the samples was estimated by the same procedures as those described in previous studies [1,2]. The synchrotron X-ray diffraction measurements at various temperatures ranging from 300 K to 25 K were conducted at the BL02B2 beamline in the SPring-8 Japan using a large Debye-Scherrer camera with a 286.5 mm camera radius [32]. The monochromatic X-ray wavelength was 0.05 nm and two-dimensional Debye-Scherrer images were detected by Imaging Plates. For measurements at low temperatures, capillaries containing ground samples were cooled using a dry $N_2$ or He gas-flow cooling device. The diffraction patterns ranging from 4 to 73° ($N_2$ gas cooling) or to 53° (He gas cooling) were obtained with a 0.01° step in 2θ, which corresponds to 0.042 nm and 0.056 nm resolution, respectively. These diffraction patterns were then subjected to the Rietveld analysis. Electrical resistivity measurements with a DC four-probe technique were conducted at 1.8–300 K using a Quantum Design Physical Properties Measurement System (PPMS) with a vibrating sample magnetometer (VSM) option at 1.9–370 K. Magnetization was measured using the same equipment under an external magnetic field up to 2 T. (More experimental information is available in online).

*Ab initio* calculations were carried out using the Density Functional Theory (DFT). We used the generalized gradient approximation (GGA) density functional by Perdew and Wang (PW91) [33] and the projected augmented waves method [34], as implemented in the computer code VASP [35]. The plane-wave basis set cutoff energy of 600 eV was used. The supercells containing 8, 16, and 32 atoms were considered and the tested meshes containing 252, 132, and 36 *k* points, respectively, were used for Brillouin zone integrations. The total energy was minimized with respect to coordinates of all atoms. For the analysis of the electronic structure, the charge-density was decomposed over the atom-centered spherical harmonics.

## 3. Result and discussion

*3.1. Electrical and magnetic measurements*

As reported previously [1], undoped samples exhibit an abrupt decrease in the resistivity at ~160 K ($T_{anom}$) with a little hysteresis around $T_{anom}$, as shown in figure 2(a), but do not exhibit the superconducting transition down to 1.8 K. The magnetization (*M*) at a fixed magnetic field (*H*) of 1 T decreased gradually with lowering temperature from room temperature, temperature dependence exhibited opposite behavior to that of conventional Pauli paramagnetism (PP), typically observed in transition element metals and FeSi [36–38]. With further lowering of the temperature from 160 to 140 K, an additional stepwise decrease was observed. However, the *M-T* dependence showed no clear anomalies, which would reflect the antiferromagnetic ordering, although the magnetic transition has been distinctly observed at ~140 K in temperature dependences of neutron diffraction, NMR, Mössbauer and specific heat [28–31]. With decreasing temperature below 140 K, the *M* value grows gradually and then starts to increase sharply at ~25 K. As shown in figure 2(b), the *M-H* curves above 25 K are almost straight whereas non-linearity starts to appear at 10 K and becomes prominent at 1.9 K (figure 2(b)). Magnetic moment at $H = 0$ can be estimated to be ~4 emu/LaFeAsO or ~$7 \times 10^{-5}$ $\mu_B$/Fe if one assumes the linear dependence of *M* on *H* in the *H* region of 1.5–2 T, but its origin remains unclear. The presence of magnetic impurity phases, such as FeAs, is a possible origin. (Contents of impurity phases estimated from the Rietveld analysis are shown in Fig. S1.)

The anomaly at ~160 K disappears in 14% F-doped LaFeAsO, which undergoes a superconducting transition at ~20 K. The *M* value becomes negative in the superconducting state due to the Meissner effect (figure 2(c)). With increasing temperature it jumps to a fairly large value at temperature just above $T_c$, and then decreases monotonically with further increase of *T*, and finally gradually increases again above room temperature, similar to undoped sample. The *M-H* curves in the normal conducting state exhibit a straight line in the high *H* region and they deviate a little down side from the line in the low *H* region (figure 2(d)). Figure 2(e) summarizes temperature-dependent magnetic susceptibilities as estimated from straight lines in figures 2(b) and (d), for LaFeAsO$_{1-x}$F$_x$ with several *x* values. Further, comparison of the curves among the superconducting samples enables us to separate the temperature-sensitive Curie-Weiss-like (CW) component from the temperature-insensitive baseline in each sample. It is noteworthy that the CW-like component increases with the F content up to 5% and then decreases with further increase in the F content. This tendency is more clearly demonstrated in figure 2(f), where the χ values are plotted against the value of *x* for three temperatures. It is, however, unlikely that this tendency is totally due to the magnetic impurity phases, such as FeAs, because the content of the dominant magnetic impurity phase of FeAs was observed to vary slightly with the F content. This tendency is not consistent with the almost constant value of $T_c$ ~26 K over the F content of 5~11% [1]. In other words, the χ value does not correlate with the generation of the superconductivity, provided that the contribution of the impurity phases to the observed χ value is small.

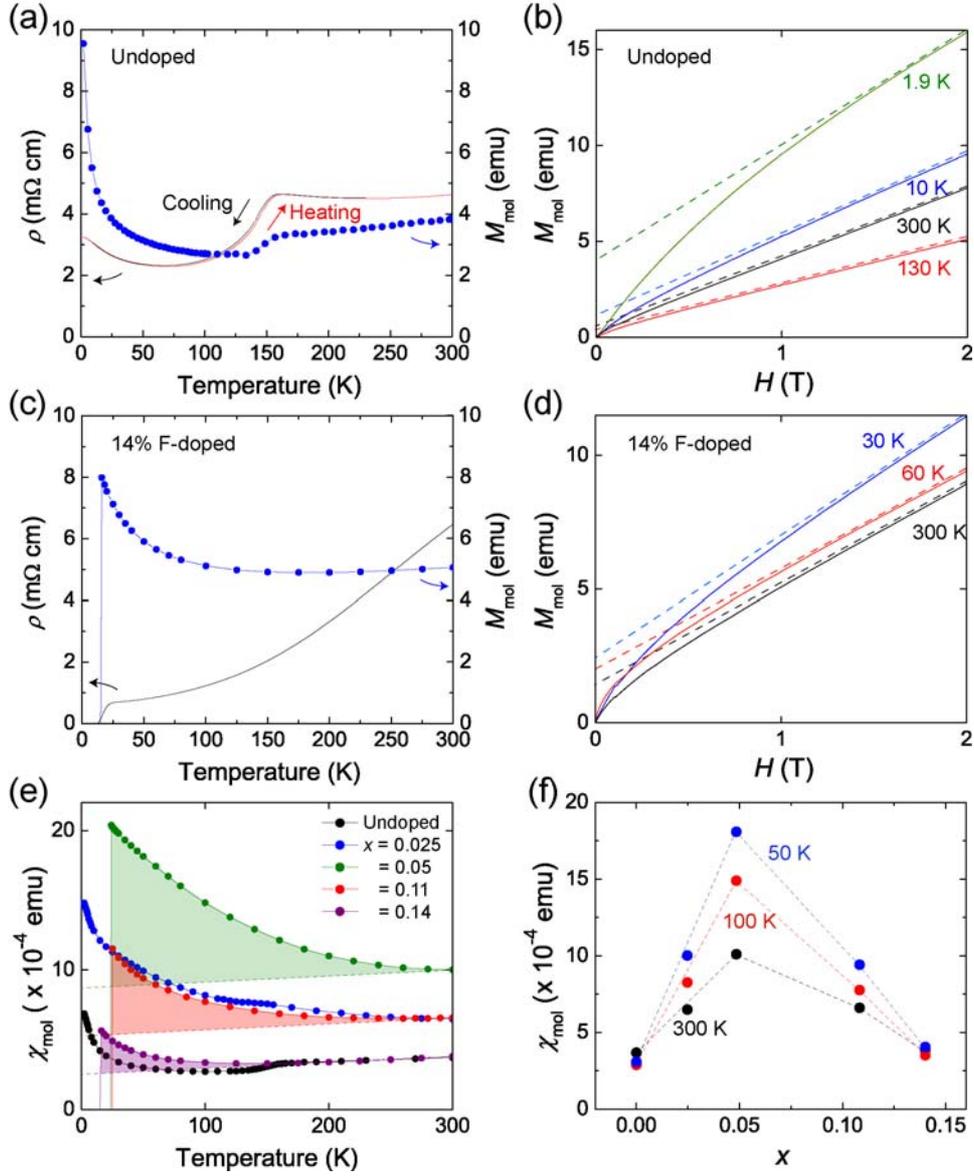

**Figure 2.** Electrical and magnetic properties of undoped and F-doped LaFeAsO. (a) Electrical resistivity (ρ) and magnetization per mol ($M_{mol}$) at a fixed magnetic field of 1 T as a function of temperature ($T$) for undoped LaFeAsO. (b) $M_{mol}$ as a function of magnetic field ($H$) for undoped LaFeAsO. (c) ρ and $M_{mol}$ at a fixed $H$ of 1 T as a function of $T$ for 14% F-doped LaFeAsO. (d) $M_{mol}$ as a function of $H$ for 14% F-doped LaFeAsO. (e) Magnetic susceptibility ($\chi_{mol}$) as a function of $T$ in various F-doped LaFeAsO samples. Dotted lines represent estimated Pauli paramagnetic components, whereas dashed areas correspond to Curie-Weiss-like components. Data are rearranged from ref. 31. (f) Magnetic susceptibility ($\chi_{mol}$) as a function of F content ($x$) in LaFeAsO for several temperatures. The midpoint $T_c$ is also shown. Data are reproduced from ref. 1.

*3.2. Crystallographic analysis*

All the diffraction peaks of the undoped sample at temperatures above the $T_{anom}$ and of the F-doped sample were assigned to the conventional ZrCuSiAs-type tetragonal crystal phase (space group of *P*4/*nmm*), except the additional weak peaks assigned to impurity phases. It is noteworthy that several peaks, including 110, 111, 112, 211 and 322 reflections of the tetragonal phase, started to split into two when the temperature was lowered below the $T_{anom}$. The peak splitting is clearly demonstrated in figure 3(a), which shows the tetragonal 322 diffraction peak profiles at various temperatures around the $T_{anom}$. On the other hand, such splitting was never observed in the diffraction peaks for the 14% F-doped sample down to 25 K.

We analyzed these XRD patterns by the Rietveld analysis to refine the crystal structures of these samples. This analysis led us to the conclusion that the anomaly is associated with the crystallographic phase transition from the tetragonal (**T**) to orthorhombic (**O**) phase. The *Cmma* space group provided the smallest $R_I$ and $R_{wp}$ value of ~2.0% and ~4.6% for the **O** phase (the fitting results are available online). The resultant structure parameters are shown in Tables 1,2. It is convenient to characterize the orthorhombic structure using a unit cell with *a* and *b* crystallographic axes rotated by 45° along the *c*-axis with respect to those of the original tetragonal cell. As a result, the number of the formula units in the orthorhombic unit cell (supercell) increases from 2 to 4, as illustrated in figure 1(b).

Cruz et al. [28] have reported the crystal phase transition of undoped LaFeAsO, but assigned monoclinic *P*112/*n* space group to the low-temperature phase. We checked our fitting results with the orthorhombic *Cmma* space group, but any peaks violating the extinction rule have not been found in the diffraction patterns down to 25 K. Structural studies of other *Ln*FeAsO structures, such as CeFeAsO, NdFeAsO and SmFeAsO [20, 39–41], have also reported that the crystal symmetry of the low-temperature phase can be successfully assigned to the *Cmma* space group, citing our report. These results support our determination of the space group for the low-temperature phase.

The lattice parameters of the undoped LaFeAsO shown in figure 3(b) as a function of *T*, confirm the existence of the crystallographic phase transition at ~160 K ($T_{anom}$). On the other hand, the F-doped samples keep the tetragonal symmetry down to 25 K, although the lattice constants become smaller with lowering the temperature. The bond lengths of the La−O, La−As and Fe−As are shown in figure 3(c) for undoped and 14% F-doped samples as a function of *T*. (The bond length of Fe−O is half of the *c* crystallographic parameter, see figure 3(b)). All the bond lengths undergo small, but abrupt changes due to the phase transition. Further, it is noteworthy that, with the F doping, the Fe−As bond length changes only slightly (smaller than 0.1%), whereas the other distances change significantly; By comparison at 120 K, the La−O distance increases by ~0.8%, on the other hand, the La−As and the Fe−O distances reduce by ~1.4% and ~0.6%, respectively. These results indicate that the F doping does not affect the geometry of the FeAs layer, in contrast to the significantly modified LaO layer. Additionally, the distance between the $(LaO)^{+\delta}$ and $(FeAs)^{-\delta}$ layers prominently decreases by F doping, suggesting that the electron doping of the FeAs layer enhances the polarization and the Coulomb interaction between the layers.

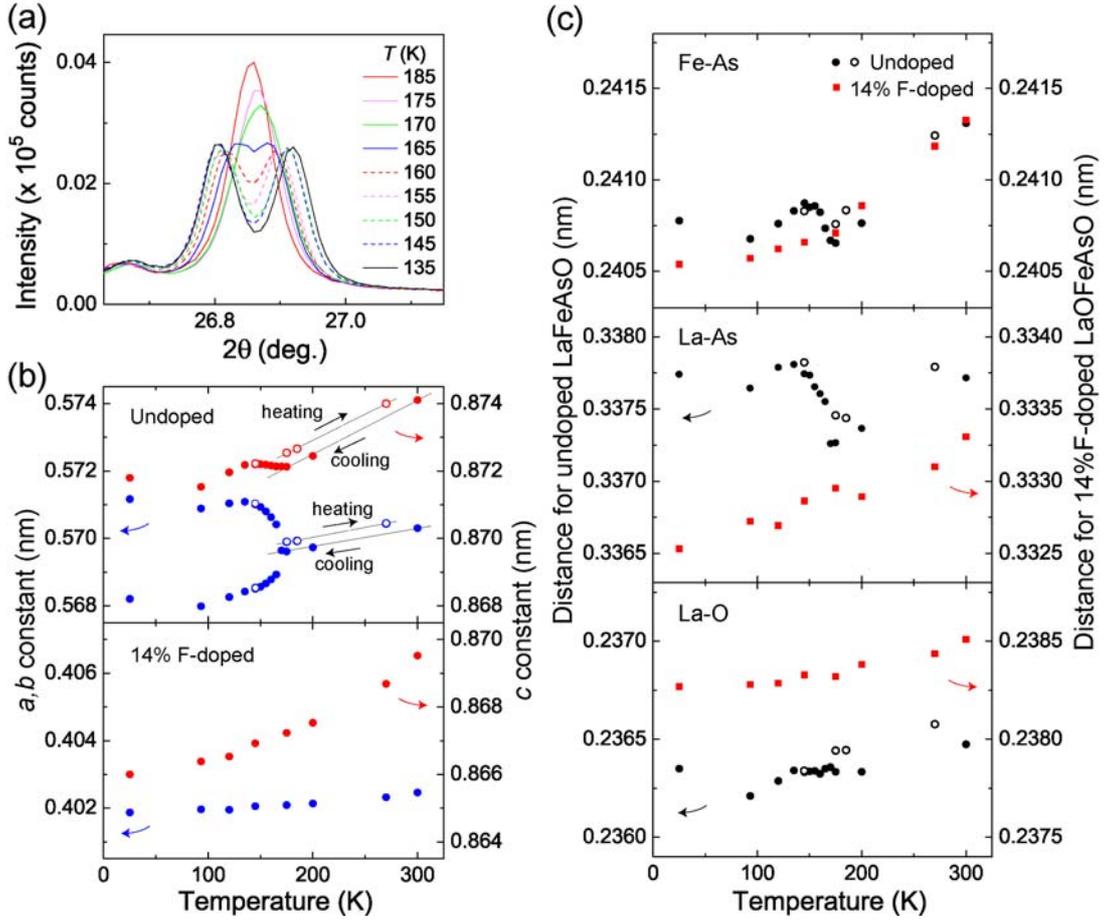

**Figure 3.** Crystallographic transition in LaFeAsO. (a) Diffraction profiles of tetragonal 322 reflections of undoped LaFeAsO for several temperatures (*T*) from 135 K to 185 K. The split peaks below 165 K are indexed as 152 and 512 reflections in the orthorhombic symmetry. (b) *a*-, *b*- and *c*-axis lengths of undoped and 14% F-doped LaFeAsO as a function of *T*. The axes of the undoped LaFeAsO are defined as the orthorhombic phase, and those in the 14% F-doped LaFeAsO are as the tetragonal one. Closed and open circles are obtained in heating and cooling process, respectively. (c) The temperature-dependent bond lengths of the La–O, La−As and Fe−As for undoped and 14% F-doped samples. Closed and open symbols in figure 3(b) and (c) represent cooling and heating processes, respectively.

**Table 1.** The structure parameters determined by Rietveld analysis for undoped LaFeAsO at (a) 300 K and (b) 120 K.

| Atom | Site | Occ. | x | y | z | B (Å$^2$) |
|---|---|---|---|---|---|---|
| (a) Undoped LaOFeAs at 300 K (P4/*nmm*, Z = 2) | | | | | | |
| a = 4.03268(1) Å   c = 8.74111(4) Å   V = 142.1524(8) Å$^3$ | | | | | | |
| La | 2c | 1.000 | 0.25 | 0.25 | 0.14134(4) | 0.476(4) |
| Fe | 2b | 1.000 | 0.75 | 0.25 | 0.5 | 0.725(15) |
| As | 2c | 1.000 | 0.25 | 0.25 | 0.65166(7) | 0.713(8) |
| O | 2a | 1.000 | 0.75 | 0.25 | 0 | 0.67(7) |
| (b) Undoped LaOFeAs at 120 K (C*mma*, Z = 4) | | | | | | |
| a = 5.68262(3) Å   b = 5.71043(3) Å   c = 8.71964(4) Å   V = 282.954(2) Å$^3$ | | | | | | |
| La | 4g | 1.000 | 0 | 0.25 | 0.14171(4) | 0.275(3) |
| Fe | 4b | 1.000 | 0.25 | 0 | 0.5 | 0.415(14) |
| As | 4g | 1.000 | 0 | 0.25 | 0.65129(7) | 0.386(8) |
| O | 4a | 1.000 | 0.25 | 0 | 0 | 0.39(6) |

**Table 2.** The structure parameters determined by Rietveld analysis for 14% F-doped LaFeAsO at (a) 300 K and (b) 120 K. The occupancy of the F atom was fixed to 14% and the isotropic atomic displacement parameter ($B$) is constrained to be that of O atom.

| Atom | Site | Occ. | x | y | z | B (Å$^2$) |
|---|---|---|---|---|---|---|
| (a) 14% F-doped LaOFeAs at 300 K (P4/*nmm*, Z = 2) | | | | | | |
| a = 4.02460(2) Å   c = 8.69525(5) Å   V = 140.8405(12) Å$^3$ | | | | | | |
| La | 2c | 1.000 | 0.25 | 0.25 | 0.14725(6) | 0.672(4) |
| Fe | 2b | 1.000 | 0.75 | 0.25 | 0.5 | 0.72(2) |
| As | 2c | 1.000 | 0.25 | 0.25 | 0.65319(10) | 0.71(1) |
| O | 2a | 0.860 | 0.75 | 0.25 | 0 | 0.91(10) |
| F | 2a | 0.140 | 0.75 | 0.25 | 0 | 0.91(10) |
| (b) 14% F-doped LaOFeAs at 120 K (P4/*nmm*, Z = 2) | | | | | | |
| a = 4.01950(2) Å   c = 8.66533(5) Å   V = 140.0004(13) Å$^3$ | | | | | | |
| La | 2c | 1.000 | 0.25 | 0.25 | 0.14774(6) | 0.422(6) |
| Fe | 2b | 1.000 | 0.75 | 0.25 | 0.5 | 0.41(2) |
| As | 2c | 1.000 | 0.25 | 0.25 | 0.65270(10) | 0.38(1) |
| O | 2a | 0.860 | 0.75 | 0.25 | 0 | 0.71(10) |
| F | 2a | 0.140 | 0.75 | 0.25 | 0 | 0.71(10) |

*3.3. Ab initio calculations*

To achieve further insight into the relation between the chemical composition, structure, magnetic properties and the superconductivity mechanism, we carried out *ab initio* calculations of both undoped and F-doped LaFeAsO using DFT and the PW91 density functional.

Our calculations suggest that the charge density distribution across the layers has $(LaO)^{+\delta}(FeAs)^{-\delta}$ character with $\delta = 0.15$ |e| and ~0.9 |e| per molecule in the PW and GTO basis sets, respectively. The density of states (DOS) near the Fermi energy (figure 5) is dominated by Fe 3*d*-states with about 10% contribution of As 4*p*-states. The total energy of the system depends on the configuration of the spins associated with Fe 3*d* electrons. For example, within the 16-atom $\sqrt{2}\times\sqrt{2}$ supercell we calculated the total energies for four non-equivalent spin configurations: ferromagnetic (FM), and three antiferromagnetic (AF1, AF2a, AF2b) ones, where AF2a and AF2b are equivalent (see figure 4). For the lattice parameters and internal coordinates of the **O** phase determined both experimentally and by calculations using the PW91 density functional, we find that $E_{FM} > E_{AF1} > E_{AF2a} = E_{AF2b}$ and relative energies of these configurations with respect to AF2a are 0.15 eV (AF1) and 0.40 eV (FM).

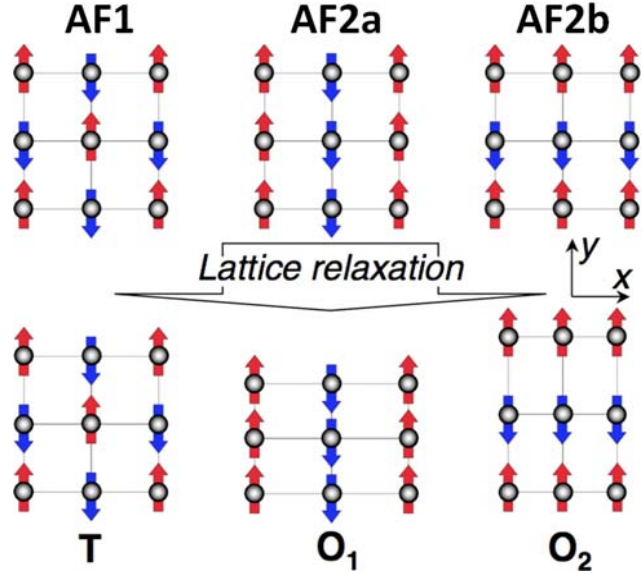

**Figure 4**. Schematic representations of Fe spin configurations in FeAs layer of undoped LaFeAsO. Arrows indicate spin-states of Fe 3*d* electrons. The upper images represent AF1 (checkerboard type) and AF2a and AF2b (spin stripe type) spin configurations in the tetragonal phase, whereas, the lower images represent the spin configurations of AF1, AF2a and AF2b in the orthorhombic phase. The *a*-axis (*x* direction) is longer than the *b*-axis (*y* direction) in $O_1$ and *vice versa* in $O_2$.

By minimizing the total energies with respect to the atomic positions as well as the lattice parameters, we find that the **O** structure has the lattice parameters of *a, b* and *c* of 0.567, 0.573 and 0.869 nm, respectively, so as $a > b$ for AF2a and $a < b$ for AF2b. In AF2, the Fe 3*d* spins along the short Fe–Fe bonds have parallel orientation and those along the long Fe–Fe bond have anti-parallel orientation. On the contrary, AF1 configuration relaxes to the **T** structure with $a = b = 0.569$ nm and $c = 0.862$ nm and remains 0.15 eV less stable than AF2. The FM configuration becomes unstable and converges to AF1. We note that relative energies of the spin-configurations are expected to depend on the accuracy of exchange-correlation functional, which needs to be investigated separately.

We analyzed the effect of the **T**–**O** transition on the electronic structure of LaFeAsO. Figure 5 shows the DOS calculated using PW91 functional near the Fermi level for the AF1 (**T**-phase) and AF2 (**O**-phase) configurations. It is clearly seen that the magnitude of the DOS near the Fermi energy is large in AF1 configuration (**T**-phase). However, it transforms into a wide depression in the AF2 configuration (**O**-phase). We note that the DOS structure is mostly determined by the spin-configuration rather than by the details of the atomic structure.

Integrating the PW91 spin density within the sphere near each Fe atom suggests that the magnetic moment per each Fe in AF2 configuration is ~1.6 $\mu_B$. The densities of spin-up and spin-down electrons are almost equivalent so as the total magnetic moment per 16-atom supercell are about $4 \times 10^{-3}$ $\mu_B$ and ~$10^{-3}$ $\mu_B$ for the tetragonal and orthorhombic structures, respectively. Although these numbers are too small to be determined accurately with the DFT, they indicate that: i) the total magnetization of FeAs layers is small, and ii) the total magnetic moment reduces during the **T**–**O** transition.

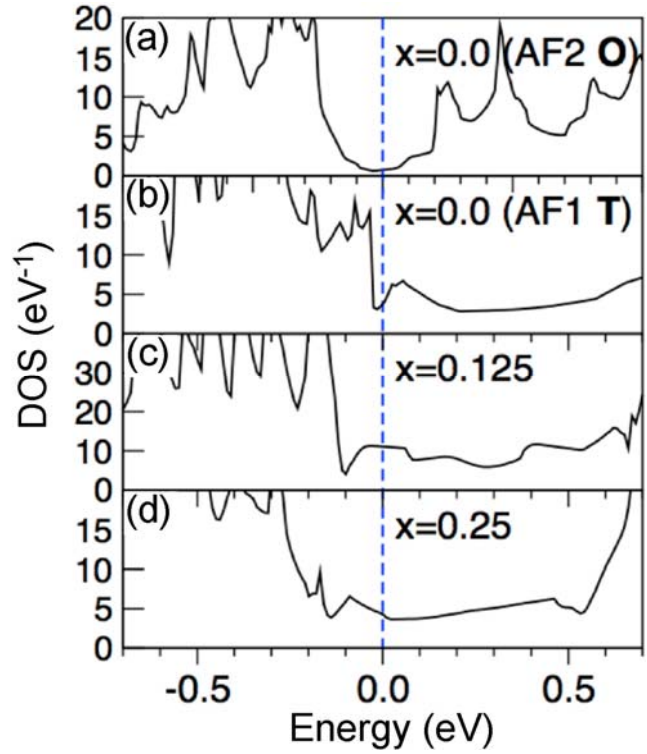

**Figure 5.** Densities of states (DOS) calculated for LaFeAsO$_{1-x}$F$_x$. (a) AF2 spin configurations in orthorhombic phase of undoped LaFeAsO ($x = 0$) (b) AF1 spin configuration in the tetragonal phase of undoped LaFeAsO ($x = 0$), (c) Paramagnetic state in the tetragonal phase for $x = 0.125$, (d) Paramagnetic state in the tetragonal phase for $x = 0.25$.

*3.4. Phenomenological model for phase transitions*

Figure 6 shows a schematic representation of potential energy surfaces (PES) calculated for undoped LaFeAsO and plotted with respect to the lattice parameters of the *Cmma* space group. The PES curves are calculated for two types of antiferromagnetic configurations, the checkerboard (AF1) and spin-stripe type (AF2), and the latter is the most stable one in agreeing with experimental observation [20]. The AF2 PES is represented by two equivalent parabolic curves with the minima at $O_1$ ($a < b$ region) and $O_2$ ($a > b$ region). The two curves split at the crossing point into upper ($AF2_U$) and lower ($AF2_L$) branches due to anharmonic electron-phonon interaction (tunneling interaction). The lower branch has the double minimum structure and, at low temperature ($T_1$ region in figure 6(b)), the system is stabilized at one of the minima. This indicates that the orthorhombic phase is stable at low temperatures. The spins on Fe atoms along short Fe–Fe bonds couple ferromagnetically in the $AF2_L$ and antiferromagnetically in the $AF2_U$, suggesting the magnetostrictive interaction plays a major role in stabilizing the orthorhombic structure. With increasing the temperature ($T_2$ region in figure 6(b)), the system migrates dynamically between the two minima due to the thermal energy, inducing the transition from the orthorhombic to tetragonal phase. The calculated double minima most likely arise from both the existence of degenerate states at the Fermi level (The energy band structure indicates two branches of Fe 3*d* orbital are quasi-degenerated at the Γ points) and the magnetostrictive interaction between the Fe spins, which can be interpreted as occurrence of cooperative Jahn-Teller effect [42,43].

In the reductively F-doped LaFeAsO$_{1-x}$F$_x$, F$^-$ ions substitute lattice O$^{2-}$ ions. This induces three main effects. First, the doping provides additional electrons to the FeAs layer and changes the calculated charge density distribution to $(LaO)^{+\delta+x}(FeAs)^{-\delta-x}$ in all considered cases ($x = 0.5$, 0.25, 0.125). The increased inter-layer ionic bonding manifests itself in the shortening of the lattice vector along the *c*-axis, observed experimentally, and in the opening of a narrow gap at ~2.5 eV below the Fermi energy separating predominantly *p*- and *d*-states of the FeAs layers.

Second, Fe magnetic moments decrease to ~1.3 $\mu_B$ for $x = 0.125$ and to below to ~0.1 $\mu_B$ for $x = 0.25$. Interestingly, we found that F doping induces strong perturbation in the spin density distribution. In particular, in the $x = 0.25$ case, the spin-down 3*d* density is localized on a single Fe atom closest to the F$^-$ impurity, while the remaining three Fe atoms share the spin-up density. However, at a more realistic doping level ($x = 0.125$), the character of the spin distribution is much more complex: the spin-density is predominantly antiferromagnetic but the local magnetic moments associated with Fe atoms become essentially disordered. Thus, we suggest that the F$^-$ doping induces strong perturbation and may destroy the antiferromagnetic spin ordering.

Finally, relaxing the lattice parameters for the F-doped systems, we find that they have tetragonal structures. For example, the relaxed 16-atom supercell ($x = 0.25$) has tetragonal lattice structure with $a = b = 0.569$ nm, and $c = 0.849$ nm. This indicates that the F-doping reduces the strength of magnetostrictive interaction and suppresses the transition into the orthorhombic phase.

The calculated DOS for AF1 state shows a pronounced depression in the DOS of both tetragonal and orthorhombic LaFeAsO at the Fermi energy. However, the F doping shifts the Fermi energy so as it is at the local DOS peak for $x = 0.125$, which agrees well with the optimal doping level of $x = 0.11$ found experimentally for this material. This DOS peak is formed predominantly by $d(xz)$ and $d(yz)$ states, which couple the interaction between the Fe atoms in the Fe sheet with the Fe–As interaction across the FeAs layer. As the doping level increases to $x = 0.25$ the value of the DOS at the Fermi energy decreases again.

Finally we note that PW91 DOS in the undoped tetragonal and orthorhombic phases of AF1 are virtually undistinguishable from each other but they are very different from those found for AF2. This suggests that the effect of the structural change due to the phase transition alone is much smaller than the effects induced by reorientation of Fe $3d$ spins. The latter are strongly affected by doping, which includes interrelated geometric, electronic and spin structure changes. In particular, the interaction between the LaO and FeAs layers increases, the Fermi level shifts to higher energies, DOS increases at the Fermi level, and antiferromagnetic order of $3d$ Fe spins may be destroyed.

F doping increases electron density within the FeAs layers and the magnetic moments associated with the Fe atoms decrease as the F doping level increases, thus, further decreasing propensity of forming the orthorhombic structure.

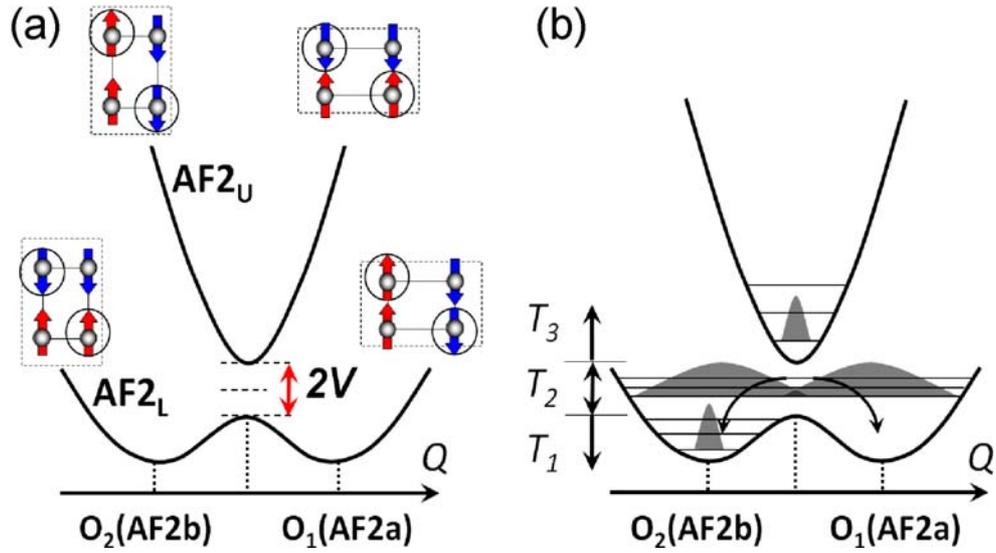

**Figure 6.** (a) Schematics of the potential energy surface calculated for AF2a and AF2b spin configurations. Their respective energy minima are in orthorhombic structures $O_1$ and $O_2$. Arrows indicate spin-states of Fe $3d$ electrons for each branch of the potential energy surface. (b) Mechanism of the structural phase transition in stoichiometric LaFeAsO. The magnetic order appears only at temperature below $T_1$.

## 4. Concluding remark

Our experimental results clearly demonstrate that the anomaly in the ρ-$T$ curve of the undoped LaFeAsO is associated with the crystallographic phase transition between the tetragonal (*P4/nmm*) to orthorhombic (*Cmma*) phases. The transition starts to occur at ~160 K, agreeing with a kink of the anomaly. The temperature is a little higher than that of the magnetic transition. The magnetic moments of Fe in the orthorhombic phase exhibit a predominantly antiferromagnetic order, forming the spin stripe type configuration in the layer. Both transitions can be explained by spin configuration-dependent potential energy surfaces derived from the *ab initio* calculations.

On the other hands, Pauli paramagnetic and spin fluctuation contribute to the magnetic susceptibility in the normal conducting state and both components are enhanced with the F-doping showing a maximum around F content of 5%. The $F^-$ ion doping at the $O^{2-}$ site suppresses the transitions and the antiferromagnetic ordering as the result of providing additional electrons confined to the FeAs layer which modifies the DOS near the Fermi energy. These results suggest that the $T_c$ and the optimal doping level in LaFeAsO and related materials can be controlled in two ways: i) the Fermi level can be shifted across the DOS peak (figure 5) by varying the dopant concentration, and ii) the DOS value at this peak can be increased by higher localization of the Fe $d(xz)$ and $d(yz)$ states. The latter can be achieved by modifying the lattice constants via doping or vacancy formation [44] in the La and/or O sublattices or replacement of La with rare earth elements with smaller ionic radii. Finally we note that the doping-induced structural modification: the overall reduction of the distance between layers and the increase in the distance between $F^-$ and $La^{3+}$ (0.249 nm) within the layer, as compared to $O^{2-}$–$La^{3+}$ distance (0.236 nm), may result in the stronger electron-phonon coupling and enhance of both spin and charge density fluctuations. Thus, the interplay between spin and charge density fluctuations [31,45] can be responsible for the high $T_c$ in LaFeAsO:F.

**Acknowledgments**

The authors thank Prof. K. Ishida of Kyoto University, Prof. H. Takahashi of Nippon University, Prof. M. Seto of Kyoto University, Dr. Y. Kobayashi of RIKEN, Prof. H. Aoki of Tokyo University, Prof. K. Kuroki of Electro-Communication University, and Prof. Y. Kubota of Osaka Prefecture University for their informative discussions. We are also grateful to J. Gavartin, C. Ruegg and G. Kresse of University College London for fruitful discussions.

# Supplementary information for "Crystallographic phase transition and high-$T_c$ superconductivity in LaFeAsO:F"

T Nomura, S W Kim, Y Kamihara, M Hirano, P V Sushko, K Kato, M Takata, A L Shluger and H Hosono

## 1. Sample preparation

We synthesized the target compounds by solid state reactions using a powder of dehydrated $La_2O_3$ (i) and a mixture powder of LaAs, $Fe_2As$, and FeAs (ii) as starting materials in a sealed silica tube. The dehydrated $La_2O_3$ power (i) was obtained by heating commercial $La_2O_3$ powders at 600°C for 10 h in air for removal of adsorbed water. To prepare the mixture powder of (ii), metal La, metal Fe, and metallic As were mixed in a ratio of 1 : 3 : 3 and put it into a silica tube in a glove box filled with a dry Ar gas. They were sealed under pumping and subsequently heated at 850°C for 10 h. Since the vapor pressure of elementary As is rather high above 600°C, the synthesis of the As-compounds as the precursor is required to avoid the possible explosion of the sealed silica ample during the main reactions. A selection of the silica tube material durable for the high temperatures and water free environment were important factors to avoid the explosion. Then, a 1 : 1 mixture of the powder of (i) and (ii) was pressed by ~10 MPa pressure to prepare pellets with 7 mm diameter × 5 mm thick or 7 mm × 10 mm × 10 mm. The pellet was placed in a silica tube, and sealed with 0.02 MPa Ar gas at room temperature. The sealed silica tube was heated at 1250°C for 40 h to synthesize undoped LaFeAsO. The Ar gas filling was effective to prevent collapse of the silica tube during this heating process. For F-doping, a part of $La_2O_3$ was replaced with a 1 : 1 mixture of $LaF_3$ and La metal in the starting materials for undoped LaFeAsO. The F concentration of the F-doped sample was estimated as ~14% from the lattice constants of samples which decrease systematically obeying the volume Vegard's rule.

## 2. XRD measurements

Optical microscope observations clarified that the sintered samples were composed of multi-grains with a maximum size of ~10 μm and some of the grains were seemingly single crystals, with an apparently orientation of the *c*-plane parallel to the sample surface. Routine X-ray diffraction measurements were performed using Bruker D8 Advance TXS with Cu Kα radiation from a rotary cathode, where the crushed powder with a particle size of ~10 μm was employed. We often observed the XRD spectra due to the orientated sample with preference of the *c*-plane parallel to the sample holder.

The high-resolution synchrotron XRD measurements at various temperatures were conducted at the BL02B2 beamline in the SPring-8 Japan. The sintered samples were ground, then packed and sealed in Lindemann glass capillaries with an inside diameter of 0.2 mm. The capillary was set at a sample stage of a large Debye-Scherrer camera with a 286.5 mm camera radius. The monochromatic X-ray wavelength was 0.05 nm and two-dimensional Debye-Scherrer images were detected by Imaging Plates. For measurements at low temperatures, the capillaries were cooled using a dry $N_2$ or He gas-flow cooling device. Temperatures of the $N_2$ and He gas were measured and calibrated them to the sample temperature. Diffraction data were collected at 300, 270, 200, 175, 145, 120 and 93 K ($N_2$ gas cooling) for 25 minutes and at 25 K (He gas cooling) for 75 minutes in each sample. In addition, the measurements of the undoped sample ranging from 185 K to 135 K by 5 K for 5 minutes were also conducted to demonstrate the detailed crystallographic changes in the phase transition. All the diffraction peaks at room temperature are assigned as those of the tetragonal LaFeAsO. Supplementary figure S1 shows ~20 times magnified XRD patterns of the undoped and 14% F-doped LaFeAsO, where the strongest diffraction peak count was ~$10^5$. There were several weak peaks which are assigned as those of impurity phases including FeAs, $La_2O_3$, and $La(OH)_3$ for the undoped sample, and FeAs, LaAs and LaOF for the F-doped sample. These impurities may not affect any significant effects on the experimental data.

To perform Rietveld analysis, the diffraction data ranging from 4 to 73° ($N_2$ gas cooling) or to 53° (He gas cooling) with a 0.01° step in 2θ were employed, which corresponds to 0.042 nm and 0.056 nm resolution, respectively. The reason why the high-angle range are restricted is that diffracted X-ray interrupted by the He gas-flow cooling devises. The weak extrinsic peaks, identified as FeAs, $La_2O_3$, and $La(OH)_3$ for the undoped sample, FeAs, LaAs, and LaOF for the F-doped sample, were also fitted as impurity phases in the analyses. Supplementary figure S2 demonstrates the fitting results of the undoped sample below (120 K) above (300 K) the crystallographic transition temperature and those at the same temperatures of the 14% F-doped samples. The $R_I$ and $R_{WP}$ values in all the fitting results are less than 3.2 and 6.4, admitting the refined model as the best possible model.

**Supplementary figures and tables**

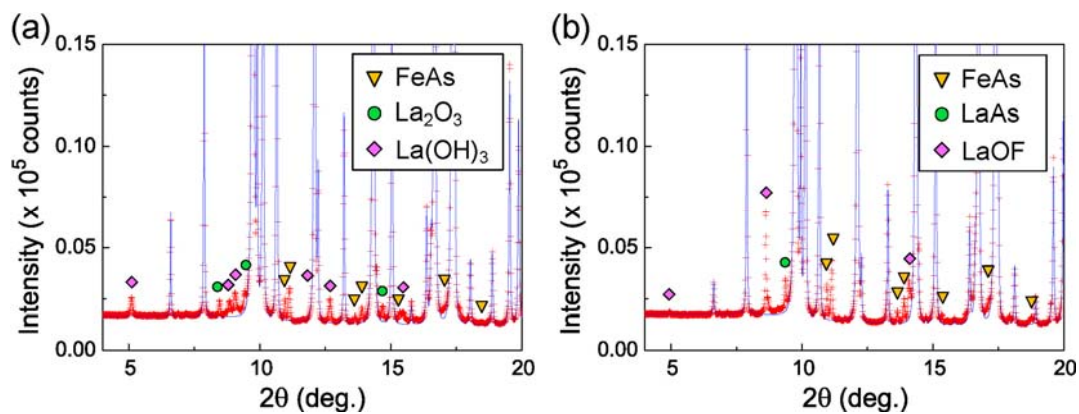

**Figure S1.** X-ray diffraction patterns in terms of undoped and 14% F-doped LaFeAsO, magnified by ~20 time in terms of the vertical axis. The amounts of impurities estimated by Rietveld analysis are ~4.7, ~0.5, ~1.6 wt% for FeAs, $La_2O_3$, $La(OH)_3$ in undoped sample, and ~3.8%, ~1.5%, ~0.9 wt% for FeAs, LaAs, LaOF in 14% F-doped sample, respectively.

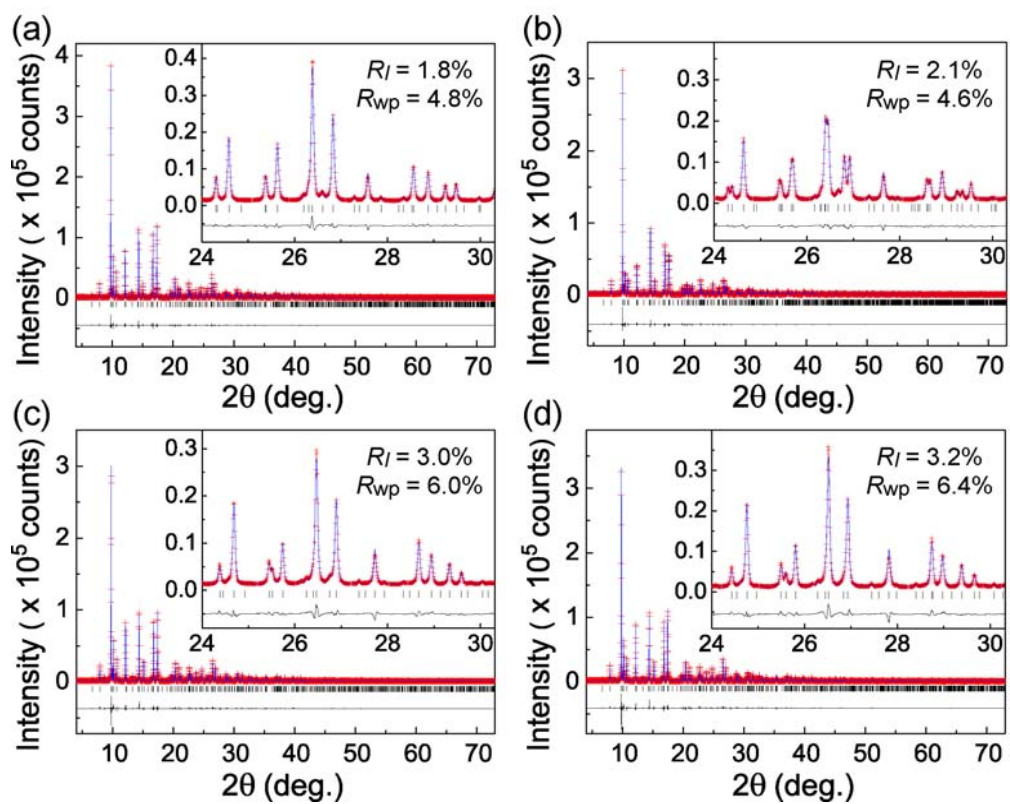

**Figure S2.** Fitting results of diffraction profiles of the undoped LaFeAsO at (a) 300 K and (b) 120 K. Those of the 14% F-doped LaFeAsO at (c) 300 K and (d) 120 K. Insets show magnifies graphs between 24° and 30°.